# FORS2 Rotating Flat Field Systematics Fixed — Recent Exchange of FORS LADC Prisms Improves the Long-known Flat-fielding Problem

Henri Boffin[1]
Sabine Moehler[1]
Wolfram Freudling[1]

[1] ESO

For many years the FORS2 instrument has suffered from artefacts in the flat fields caused by surface inhomogeneities on the longitudinal atmospheric dispersion corrector (LADC) that affected high-precision photometric and spectroscopic measurements. Recently, the FORS LADC prisms were exchanged, and our analysis of a large number of flat fields shows that this exchange has resulted in a significant decrease in the level of small-scale artefacts.

One of the most recent techniques used to probe the atmospheres of exoplanets is transmission spectroscopy, which relies on measuring how the transit depth of a planet crossing the disc of its host star varies with wavelength. Such measurements require extreme precision — at the level of 100 ppm — and can only be done with space- or large ground-based telescopes. After some pioneering work done by Bean et al. (2010) with the FORS2 instrument attached to the Very Large Telescope (VLT), this field was brought to a halt at ESO until very recently because FORS2 measurements produced unexpectedly high systematics in the differential light curves obtained with this instrument. These systematics turned out to be impossible to calibrate out reliably (see Boffin et al. [2015] for a more extended discussion).

In parallel, Moehler et al. (2010) studied the twilight flat fields obtained with the two FORS instruments and found structures that rotated with the field rotator, structures that had a significant impact on photometric measurements. These authors concluded that the origin of these structures was the LADC and developed a technique to minimise the impact of these artefacts on photometry. The current FORS2 calibration plan routinely acquires the necessary data for the implementation of this method (Bramich et al., 2012; Coccato et al., 2014). These structures were also at the origin of the systematics affecting exoplanet transmission spectroscopy.

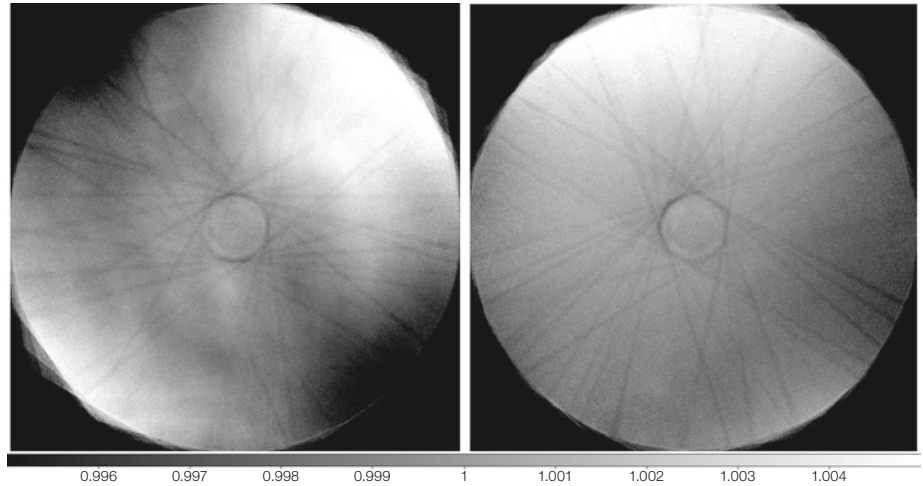

Figure 1. Comparison of rotator-angle dependent features in the FORS2 flat fields (i.e., corrected for detector/filter dependent features, see text for details) for the R_SPECIAL filter, before (left) and after (right) the prism exchange of the LADC. The scale represents artefacts on a scale between 0.995 (black) to 1.005 (white).

In order to improve the spectroscopic performance, the FORS2 LADC prisms were exchanged with the uncoated ones of FORS1 — which has been dismounted to leave space for other, second generation instruments at the VLT. The preliminary results were clearly very encouraging, as the red noise due to systematics significantly decreased after the exchange, making it possible to perform transmission spectroscopy with FORS2 again (Sedaghati et al., 2015).

As part of the regular calibration plan, sufficient data have been collected since the exchange of the prisms to allow us to look again at the level of rotating structures in the twilight flat fields of FORS2 and investigate whether they have largely disappeared. To this aim, we collected FORS2 twilight flat fields taken with the standard setup with the filters b_HIGH, v_HIGH, R_SPECIAL, and I_BESS. As part of the mitigation process, FORS2 twilight flat fields are now taken almost every four to seven nights (in the above filters at least) with a random rotator angle on sky.

We selected at most five frames per 5° angle range for rotator angles between ±180° (angles outside that range were adjusted by adding/subtracting 360°). We then combined the frames from the two detectors, after correcting each by its overscan bias, normalised each frame by its median and then median-stacked all the frames (per filter and detector) and normalised each frame by the corresponding stack to remove any rotator-independent structure. Each normalised frame was then rotated to a rotator angle of 0° and to a random rotator angle. Finally all rotator angle 0° and random rotator angle frames were again median-stacked.

This exercise was performed for frames observed between 1 February and 31 October 2014 (before the LADC prism exchange) and between 13 November 2014 and 4 August 2015 (after the LADC prism exchange). Figure 1 shows the result for filter R_SPECIAL. Note that the dark stripes come from the gap between the two detectors.

The frames are displayed in the contrast range of 0.995–1.005, i.e. ±0.5 %. The small-scale structures clearly visible in the old data are gone, leaving only a gradient across the field. This probably explains the improvement we have seen in the precision with which it is possible to perform transmission spectroscopy. The presence of the gradient indicates that photometry over the whole field still requires attention to the rotator angle and, therefore, that the current calibration plan should not be changed, as we still need to collect the data to be able to correct for rotation effects.

We will continue to monitor the structures in the twilight flat fields, to prevent users trying to perform observations at the level that is not attainable should the optics in FORS2 be degraded. As there is no longer a coating on the FORS2 LADC prisms, we do not expect, however, to encounter the same problem as before, which was due to a degradation of the anti-reflection coating. The result of our monitoring will be put on the ESO FORS2 public web pages[1].

#### Acknowledgements

We would like to thank Jonathan Smoker for a careful reading of the manuscript.

#### Links

[1] FORS2 News page: http://www.eso.org/sci/facilities/paranal/instruments/fors/news.html

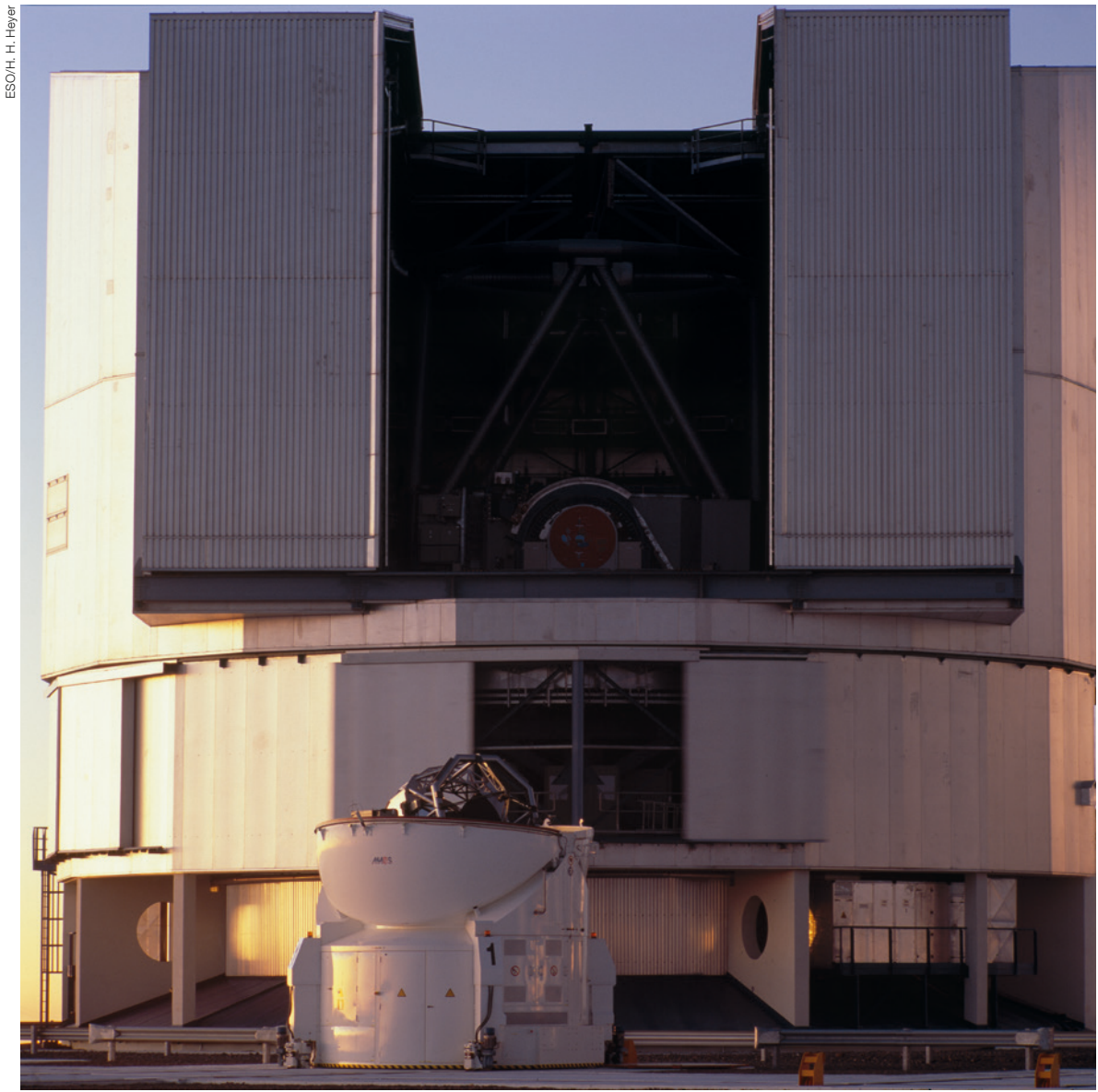

Unit Telescope 1 (Antu), the current location of the FORS2 instrument. In this photograph (from 2007), one of the Auxiliary Telescopes, belonging to the Very Large Telescope Interferometer (VLTI), is positioned in the foreground.